\documentclass{article}
\usepackage{spconf,amsmath,graphicx}
\usepackage{booktabs}
\usepackage{enumitem}
\usepackage{cite}
\usepackage{comment}
\usepackage{color}

\title{Leveraging Redundancy in multiple audio signals \\ for Far-Field Speech Recognition}
\name{\begin{tabular}{@{}c@{}}
     Feng-Ju Chang,
     Anastasios Alexandridis,
     Rupak Vignesh Swaminathan,
     Martin Radfar, \\
     Harish Mallidi,
     Maurizio Omologo, 
     Athanasios Mouchtaris, 
     Brian King, 
     Roland Maas
\end{tabular}}
\address{Alexa Speech, Amazon, USA}
\begin{document}
\ninept
\maketitle
\begin{abstract}
To achieve robust far-field automatic speech recognition (ASR), existing techniques typically employ an acoustic front end (AFE) cascaded with a neural transducer (NT) ASR model. The AFE output, however, could be unreliable, as the beamforming output in AFE is steered to a wrong direction. A promising way to address this issue is to exploit the microphone signals before the beamforming stage and after the acoustic echo cancellation (post-AEC) in AFE. We argue that both, post-AEC and AFE outputs, are complementary and it is possible to leverage the \textit{redundancy} between these signals to compensate for potential AFE processing errors. We present two fusion networks to explore this redundancy and aggregate these multi-channel (MC) signals: (1) Frequency-LSTM based, and (2) Convolutional Neural Network based fusion networks. We augment the MC fusion networks to a conformer transducer model and train it in an end-to-end fashion. Our experimental results on commercial virtual assistant tasks demonstrate that using the AFE output and two post-AEC signals with fusion networks offers up to 25.9\% word error rate (WER) relative improvement over the model using the AFE output only, at the cost of $\leq$ 2\% parameter increase.

\end{abstract}

\section{Introduction}

Neural transducer (NT) based models, such as the recurrent neural network transducer (RNN-T)~\cite{graves2012sequence}, the conformer transducer (CT)~\cite{li2021better,gulati2020conformer}, and the ConvRNN-T~\cite{radfarconvrnn} have achieved state-of-the-art results in single-channel ASR. In many cases, voice assistants are far from the sound source (such as in smart speakers) and operate in noisy conditions (noise from various sound sources or multiple speakers) which makes the acoustic scenario challenging. For such cases, voice assistants typically employ multi-microphone or microphone-array based acoustic front ends (AFEs) that use beamforming and adaptive noise reduction techniques
~\cite{haeb2019speech,omologo2001speech,wolfel2009distant,kumatani2012microphone,kinoshita2016summary,menne2016rwth,haeb2020far} to provide an enhanced signal for ASR.

Typically, the AFE is optimized based on a signal-to-noise ratio (SNR) objective under the knowledge of the microphone array geometry and specific noise field assumptions~\cite{wolfel2009distant}. The resulting enhanced signal could thus be suboptimal for the ASR task, as the optimization does not consider the ASR objective. Moreover, undesired beam switches caused by interfering sources and strong assumptions about the noise field could degrade the performance of the AFE and thus the downstream ASR task.
Recently, beamformers~\cite{ochiai2022mask,heymann2016neural,erdogan2016improved} built on neural networks have relaxed the above constraints and delivered superior results compared to AFEs based on signal processing methods. Moreover, the neural beamformer can directly be optimized together with the ASR objective in an end-to-end manner~\cite{Wu2019,ochiai2017multichannel,chang2019mimo,chang2020end,kumatani2019multi,li2016neural,xiao2016deep,meng2017deep,liu2014using}. 

\begin{figure}[t]
\centering
\includegraphics[scale=0.27]{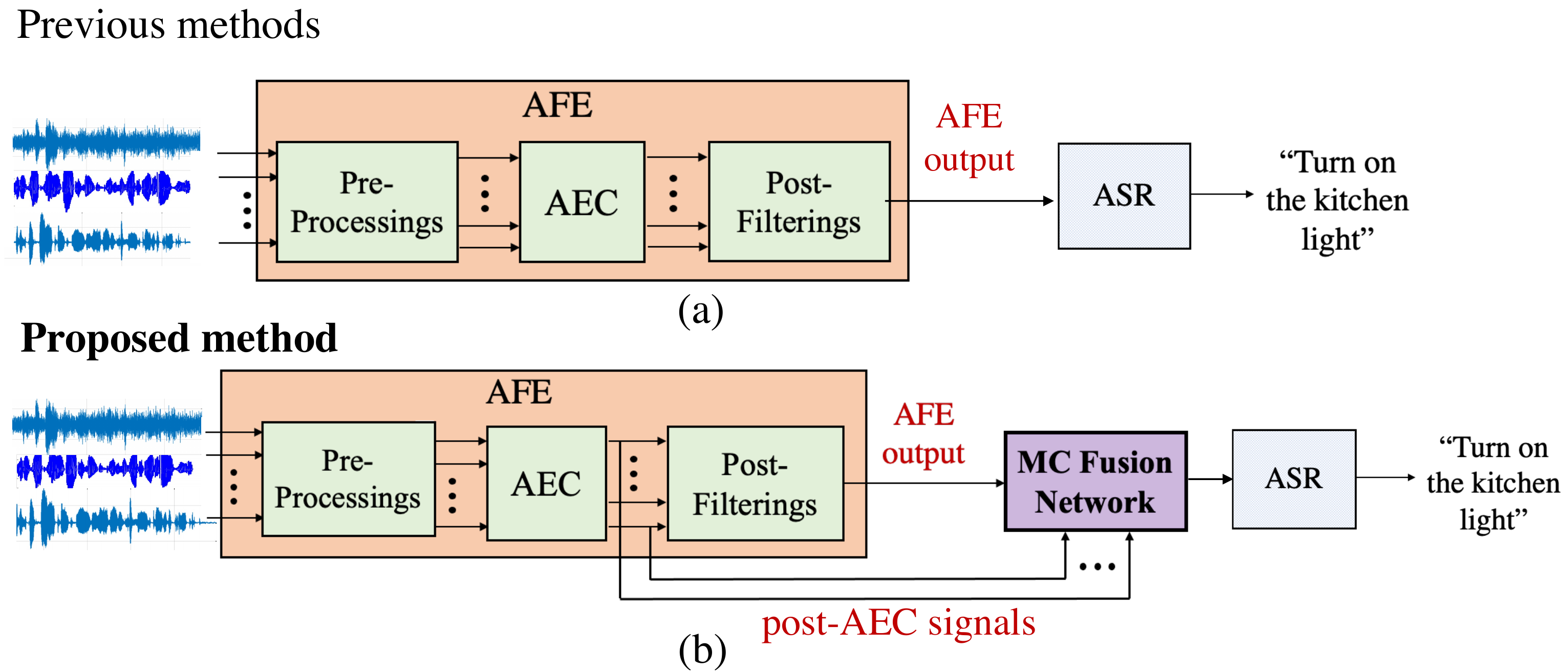}
\vspace{-2.5mm}
\caption{A high level diagram of the proposed method. Different from previous methods~\cite{li2016neural,xiao2016deep,meng2017deep,liu2014using,Wu2019} (a), we leverage the redundancy between the AFE output and post-AEC signals via a MC fusion network (b), before feeding the signal to the ASR model.}
\label{fig:mc-fusion-net}
\end{figure}

This work is motivated by the opportunity of best exploiting the information generated by an AFE~\cite{ayrapetian2021asynchronous}, which also includes a multichannel AEC (MCAEC) processing aimed at suppressing playback or other known interference signals. AEC outputs can be complementary to the beamformed signal, in particular under noisy conditions. Thanks to this feature, we design \emph{fusion nets} that utilize the AFE output and post-AEC signals, and produce a hidden representation that is forwarded to the ASR component, as illustrated in Fig.~\ref{fig:mc-fusion-net}\footnote{The presented AFE block is a simplified version of~\cite{ayrapetian2021asynchronous}. The AFE output indicates the result of a combination of MCAEC, beamforming, and noise reduction algorithms.}. In this way, the ASR model can be trained in an end-to-end fashion, by exploiting such complementarity. A similar approach was proposed in~\cite{chang2021end, chang2021multi}. However, in~\cite{chang2021end, chang2021multi} all input channels are directly forwarded to the ASR audio encoder, which makes it computationally more expensive and less attractive for streaming ASR.


Two \emph{fusion nets} are based on (a) the frequency-LSTM~\cite{FLSTM} with a neural beamformer~\cite{Wu2019} pre-processing the post-AEC signals (NBF+F-LSTM), and (b) a convolutional neural network (CNN). In the NBF+FLSTM fusion network, we expand the neural beamforming approach of~\cite{Wu2019} (that was initially proposed for DNN-HMM ASR acoustic modeling) and utilize an F-LSTM to produce the final hidden representation based on the multiple signals. Note that in~\cite{Wu2019}, the authors used pre-AEC microphone signals rather the post-AEC signals. The CNN fusion network, on the other hand, is inspired by~\cite{radfarconvrnn,yeh2019transformer,mohamed2019transformers}, but here we use convolutions to model multi-channel signals. 
Our fusion networks are used to augment a CT ASR model~\cite{li2021better,gulati2020conformer}. Note that all of our models are trained end-to-end, are streamable, while the fusion network adds less than $2\%$ of additional parameters in the model architecture. Our experiments show that the proposed methods outperform the CT with only the AFE output, offering up to 25.9\% WER relative improvement. 

\section{Neural transducers for ASR}
A shared component between the baseline system (Fig.\ref{fig:mc-fusion-net}(a)) and our method (Fig.\ref{fig:mc-fusion-net}(b)) is the NT ASR. It typically consists of an encoder network, a prediction network and a joint network. The encoder network encodes the input time frames $\mathbf{x_t}$ into high level representations $h_t^{enc}$. The prediction network uses the previously predicted word-pieces $\mathbf{y_{u-1}}$ and produces the intermediate representations $h_u^{pred}$. The joint network combines the representations of $h_t^{enc}$ and $h_u^{pred}$ by first applying a joint operation and passing the output through a series of dense layers with activations and a final softmax that gives the probability distribution of word-pieces.
The neural transducer is trained based on the RNN-T loss using the forward-backward algorithm~\cite{graves2012sequence} that accounts for all possible alignments between the acoustic frames and the word-pieces of the transcription. 
The architecture of the encoder and prediction network can vary, with typical choices being RNN layers~\cite{graves2012sequence}, transformer~\cite{dong2018speech} or conformer blocks~\cite{li2021better,gulati2020conformer}. In this work, we use a conformer encoder and an LSTM decoder (CT). Note that our fusion networks can be applied with any NT, and not limited to CT.

\section{Proposed MC fusion networks}
\label{sec:mc-fuion-net}

\begin{figure}[t]
    \centering
    \includegraphics[scale=0.3]{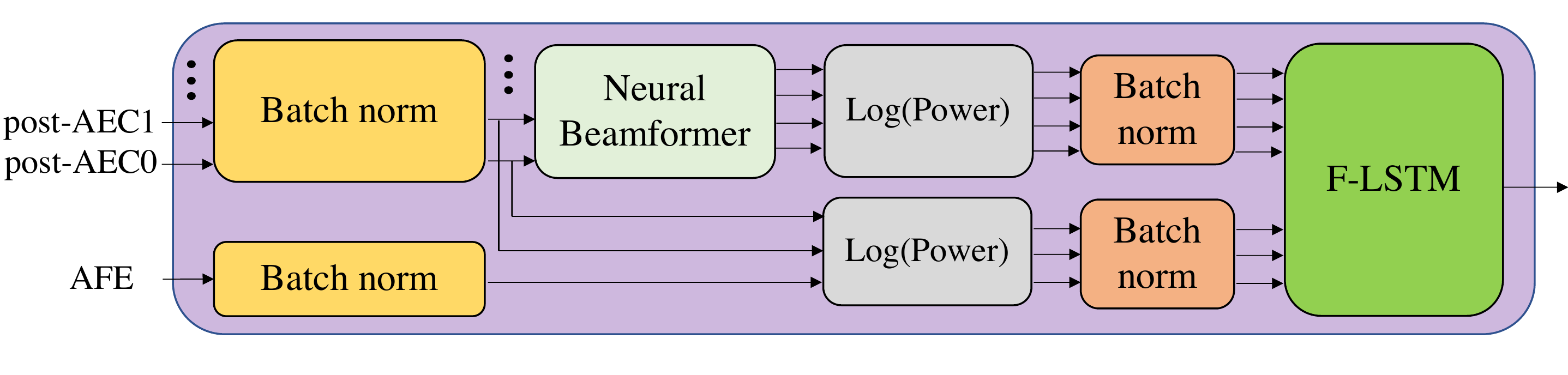}
    \caption{The neural beamformer+Frequency-LSTM (NBF+F-LSTM) fusion network. Note that the same colored batch norm layers indicates the shared parameters.}
    \label{fig:bf_flstm_diagram}
\end{figure}

Our goal is to empower the CT ASR model to leverage multiple signals processed by different techniques and explore the redundancy, including (1) An enhanced signal produced by a signal processing based AFE~\cite{ayrapetian2021asynchronous} that performs MCAEC, adaptive beamforming and noise reduction, and (2) The signals after AEC (post-AEC signals). These signals are processed by the MC fusion network (Figure~\ref{fig:mc-fusion-net}) leading to a high level representation that is input to the encoder network of the CT model. We propose two MC fusion networks. 



\subsection{Neural beamformer (NBF)+F-LSTM Fusion Network}
This approach utilizes a trainable neural beamformer inspired from~\cite{Wu2019} where authors used neural beamforming techniques for DNN-HMM ASR acoustic modeling. 
The approach is comprised of two major steps: a neural beamformer to perform spatial filtering on the multichannel inputs and a Frequency LSTM (F-LSTM)~\cite{FLSTM} layer to combine the AFE output and post-AEC signals. The architecture of the fusion network is shown in Figure~\ref{fig:bf_flstm_diagram}.
First, a batch normalization is applied on the frequency bins of the input signals to stabilize the training and make convergence faster.

\noindent \textbf{Neural beamforming network.} 
The $M$ batch normalized post-AEC signals $\mathbf{X}(t,\omega)$ enter the neural beamforming block, where a filtering operation is performed as: 
\begin{equation}
\label{eq:beamforming}
Y(t,\omega, \mathbf{p}) = \mathbf{w}^H(t,\omega,\mathbf{p}) \mathbf{X}(t,\omega)    
\end{equation} 
where $H$ denotes the Hermitian operation, $\mathbf{w}(t, \omega, \mathbf{p})$ is a complex weight vector for beamforming at position $\mathbf{p}$; $t$ and $\omega$ denote the time frame and angular frequency. 

\begin{figure}[t]
\centering
\includegraphics[scale=0.3]{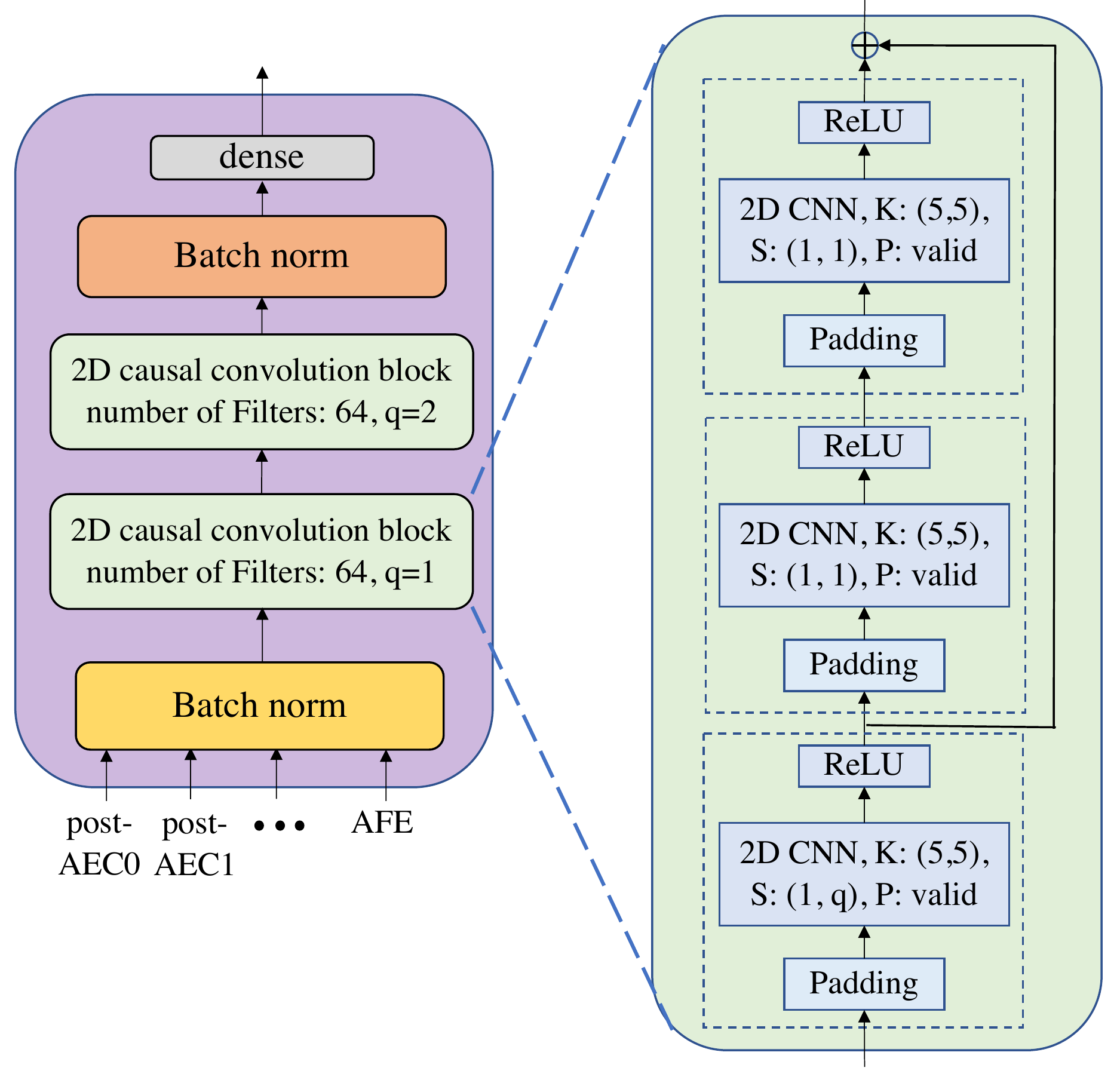}
\vspace{-0.5cm}
\caption{The Convolutional (CNN) based fusion network}
\label{fig:conv-mcf}
\end{figure}


The complex vector  multiplication in Eq.~(\ref{eq:beamforming}) for a given frequency $\omega$ can be expressed in matrix form as:
\begin{equation}
\begin{bmatrix}
Re(Y) \\ Im(Y)
\end{bmatrix} = 
\begin{bmatrix}
Re(w_0) & Im(w_0) \\
-Im(w_0) & Re(w_0) \\
\vdots & \vdots \\
Re(w_{M-1}) & Im(w_{M-1}) \\
-Im(w_{M-1}) & Re(w_{M-1}) 
\end{bmatrix}^T
\begin{bmatrix}
Re(X_0) \\
Im(X_0) \\
\vdots \\
Re(X_{M-1}) \\
Im(X_{M-1}) 
\end{bmatrix}
\end{equation}
where  $(t, \omega, \mathbf{p})$ have been omitted for clarity and $w_i$ denotes the i-th element of the complex beamforming weight vector $\mathbf{w}$. Given this formulation, the beamforming operation can be incorporated into the neural network by using $\Omega$ $2 \times 2M$ matrices (one for each frequency bin), where $\Omega$ is the total number of frequency bins. These matrices represent the beamformer weights.
Note that the complex signal features are treated as two real-valued inputs; both the real and imaginary parts of the complex neural beamformer weights are learnable. 

Initially, the weights are initialized to steer the beamformer into $n_l$ look directions that are uniformly distributed around the array. During training, the $n_l$ sets of weights are optimized according to the RNN-T loss function. The neural beamformer produces $n_l$ signals in $n_l$ look directions.  
 Next, all neural beamformed signals are considered together with the original microphone array signals and the AFE output. These are a total number of $n_l + M + 1$ signals. For each signal, we compute its power for each frequency bin,  by applying a square and sum operator to the real and imaginary parts. The powers then go through a ReLU activation function and a \textit{log} function (that also operate on the frequency axis), in order to get an estimate of the log power spectrum. Finally, we add another batch normalization layer operating on the frequency axis of each signal.

\noindent \textbf{Fusion using Frequency LSTMs}.
Finally, the $n_l + M + 1$ signals are fused into a single hidden representation. For that, we apply an F-LSTM layer~\cite{FLSTM}. An F-LSTM is a regular LSTM layer that operates on the frequency axis instead of the time axis. 
To prepare the input for the F-LSTM, we concatenate all signals together in one vector and take chunks along the frequency axis with a certain window $w$ and stride $s$ in order to create sequences in the frequency domain of length $w$. This input is then passed through an LSTM. The output of the LSTM is projected to a hidden representation of $d_{flstm}$ that is the input to the audio encoder of the CT.

\subsection{Convolutional Neural Network (CNN) Fusion Network}
\label{sec:conv}

Figure \ref{fig:conv-mcf} illustrates the Convolutional Neural Network (CNN) based MC fusion network. Its design is motivated by ConvRNN-T~\cite{radfarconvrnn} with several adaptations. First, the input channels contain the AFE output and post-AEC signals. They are concatenated together to form a tensor of $T \times \Omega \times C$, where $T$ is the number of time-frames, $\Omega$ is the number of frequency bins, and $C=M+1$ is the number of input channels.
The tensor is mean-normalized by the batch norm layer over the frequency axis before feeding into the two convolutional blocks. The preliminary studies show that this batch norm is critical to stabilize the training. Each convolutional block has three convolutional layers and we use two blocks leading to 6 convolutional layers in total, as shown in Figure \ref{fig:conv-mcf}. Each convolutional layer is followed by a ReLU activation, and an additional residual link is added from the first ReLU input with the output. To ensure the convolution computations are causal (since our model is streamable), we zero-pad the time dimension from the left with kernel size ($K$) minus one zeros. Note that the second convolutional block downsamples the frequency bins by two (stride size $q=2$) to maintain the model size. After the convolutional blocks, the second batch norm normalizes both the mean and scale of CNN outputs along the frequency axis, which we found offering better convergence in our studies. The final dense layer then projects the normalized outputs to the $d_{cnn}$ dimension, and feed it to the CT audio encoder.

\section{Experiments}
\label{exp_setup}


\subsection{Datasets}
To evaluate the proposed methods, we conduct a series of ASR experiments using our in-house de-identified far-field datasets
\footnote{To the best of our knowledge, no dataset exists that contains both AFE outputs and post-AEC microphone array signals.}.
The datasets are for the English language and virtual assistant tasks. The device-directed speech data was captured using a smart speaker with 7 microphones, and a 63 mm aperture.
The training sets contain $\sim 500K$ hours of single-channel (SC) data and $\sim 60K$ hours of multi-channel (MC) data. The SC data consists of the AFE output~\cite{ayrapetian2021asynchronous}; the AFE takes as input the 7 microphone signals. The MC data contains 2 post-AEC signals of aperture distance in addition to the AFE output. When training/testing the MC models on the SC data (which only contains the AFE output), we simply replace the 2 post-AEC signals with zeros.

We use the following SC and MC test sets: (1) \textit{Head}: 10 {hours} of SC data that represent the head distribution of a far-field voice assistant task. 
(2) \textit{MC clean}: 3 {hours} of MC data of high SNR for a virtual assistant task. (3) \textit{MC noisy}: 20 {hours} of MC data for a virtual assistant task collected in a controlled condition. This dataset contains more challenging acoustic scenarios covering two common device locations and three noise conditions. The Device locations include (1) \textbf{DevLoc1}: The device is right in front of the speaker (5 feet away). (2) \textbf{DevLoc2}: The device is in front of the speaker but far away in the corner (15 feet away). The noise conditions are (1) \textbf{Silence}: no background noise (2) \textbf{Appliance}: appliance noises, e.g., air conditioner, and (3) \textbf{Media}: The noises from the TV, radio, etc.


\subsection{Model Configurations}
\subsubsection{Baseline methods}
We use the conformer transducer (CT) as the ASR component for all the following baseline models. (1) \textbf{AFE CT}: It serves as the baseline using only the AFE output and no fusion net.
(2) \textbf{CA CT}: This baseline uses the channel attention (CA) based fusion network as proposed in \cite{chang2021multi}. We feed both AFE output and post-AECs to this model for evaluating our fusion networks. To maintain the comparable model size, we employ one cross-channel attention layer and select the best number of attention heads (among 1, 2, 4) and head sizes (among 512, 256, 128) based on the validation set performances. All CTs in the baselines share the same configurations: Two convolutional layers followed by 14 conformer blocks. The convolutional layers have 128 kernels of size 3, the stride of 2 and 1 along the time axis for the first and second convolution respectively. Each conformer block contains the FeedForwardNetwork (FFN) module with 1024 units, the convolutional module with kernel size 15, and the attention module with the embedding size 512, and 8 heads. The number of parameters is 81.44 million. The CT encoder outputs are projected to 512-dimensional hidden representations, before being fed into the joint network. The prediction network consists of two LSTM layers with 1024 units. The output is also projected to a 512-dimension to match the audio encoder output.  The joint network consists of a FFN with 512 units.
We used the Adam optimizer \cite{kingma2014adam} and varied the learning rate following \cite{vaswani2017attention,dong2018speech}.

\subsubsection{Fusion Networks}
\label{fusion_net_configs}
\noindent \textbf{Neural beamformer (NBF) + F-LSTM}:  For the neural beamformer, we set the number of look directions to $n_l = 4$. Our preliminary analysis showed that increasing look directions does not lead to noticeable improvements.
For the F-LSTM layer we use a window of 48 frequency bins with a stride of 15 and employ two bi-directional LSTMs with 16 units. The outputs of the F-LSTM is projected to $d_{flstm}=192$-dimensional vector in match the feature size of baselines. The total number of parameters is 372K, which takes only $\sim 0.5\%$ of the parameter size of the conformer encoder.
 
 \noindent \textbf{Convolutional Fusion Network (CNN)}: It contains two 2D causal convolution blocks, each of which has three convolutional layers. The first 2D causal convolution block has 64 kernels of size $K$=(5,5), the stride size equals to 1 along both the time axis and the stride size along frequency axis, $q$=1. The second 2D causal convolutional block has the same configuration as the first one but with the stride size along the frequency axis, $q$=2. The final dense layer projects the normalized CNN block outputs to the $d_{cnn}=192$-dimensional vector. The total number of parameters is 1.4M, which takes only $\sim 1.8\%$ of the parameter size of the conformer encoder. All the models are of comparable size, $\sim 82$ million parameters in total. 

\subsection{Input and Output Configurations}
The input features for AFE CT model are Log-filter bank energies (LFBEs) and are extracted with a frame rate of 10 ms with a window size of 32 ms from audio samples sampled at $16$~kHz. The features of each frame are stacked with the ones of left two frames, followed by the downsampling by a factor of 3 to achieve low frame rate, resulting in 192 feature dimensions. The proposed methods use the Short Time Fourier Transform (STFT) features, which are computed with a 512-point Fourier Transform on segments using a 32 ms window with a frame rate of 10 ms. 
We separate the complex values into their real and imaginary parts, resulting in 512-dimensional features for every frame. Similarly, we use a left context of 3 frames.
A subword tokenizer \cite{sennrich-etal-2016-neural} is used to create output tokens from the transcriptions; we use 4K tokens.

\section{Results and Discussions}

\begin{table}[t]
    \caption{WERRs(\%) of the proposed models and channel attention (CA) based fusion networks \textbf{over AFE CT}, on the \textit{HEAD}, \textit{MC clean}, and \textit{MC noisy} test sets. All three models use the AFE output and post-AECs as inputs. A higher number indicates a better WER.}.
    \label{tab:werrs_main}
    \centering
    \resizebox{0.99\linewidth}{!}{
    \begin{tabular}{r|c|c|c}
    \toprule
        $\text{Fusion Network} (\downarrow) / \text{Testset} (\rightarrow)$ & \textit{HEAD} & \textit{MC Clean} & \textit{MC Noisy}  \\ \midrule
       NBF+F-LSTM & 5.66 & 5.31 & \textbf{10.75} \\ 
       CNN & \textbf{7.76} & \textbf{6.10} & 10.38 \\
       CA~\cite{chang2021multi} & 4.61 & 3.74 & 8.96
    \end{tabular}}
\end{table}

\begin{table}[t]
    \caption{Ablation study given WERRs(\%) \textbf{over AFE CT}: Comparisons of the proposed models to the ones removing either the fusion network or post-AECs, on the \textit{HEAD}, \textit{MC clean}, and \textit{MC noisy} test sets. A higher number indicates a better WER.}.
    \vspace{-0.2cm}
    \label{tab:werrs_ablation}
    \centering
    \resizebox{0.99\linewidth}{!}{
    \begin{tabular}{r|c|c|c}
    \toprule
        $\text{Model} (\downarrow) / \text{Testset} (\rightarrow)$ & \textit{HEAD} & \textit{MC Clean} & \textit{MC Noisy}  \\ \midrule
       AFE CT & baseline & baseline & baseline \\ \hline
       \textbf{+ post-AECs+NBF+FLSTM (Ours)} & \textbf{5.66} & \textbf{5.31} & \textbf{10.75} \\
       + post-AECs (\textbf{no fusion net}) & 1.89 & 3.54 & 7.92 \\
       + NBF+FLSTM (\textbf{no post-AECs}) & 0.84 & 1.18 & 6.70 \\ \hline
        \textbf{+ post-AECs+NBF+CNN (Ours)} & \textbf{7.76} & 6.10 & \textbf{10.38} \\
       + post-AECs (\textbf{no fusion net}) & 1.89 & 3.54 & 7.92 \\
       + CNN (\textbf{no post-AECs}) & 5.24 & \textbf{8.66} & 6.89 \\ 
    \end{tabular}}
\end{table}

We report the relative word error rate reduction (WERR) of the proposed methods against baselines throughout the experiments\footnote{The absolute WERs for the baselines are under 10\%. The absolute WERs are not reported here due to institution policy.}.
Given a model A's WER ($\text{WER}_A$) and a baseline B's WER ($\text{WER}_B$), the WERR of A over B can be computed by $(\text{WER}_B - \text{WER}_A) / \text{WER}_B$; a higher WERR value indicates a better performance.

Table~\ref{tab:werrs_main} presents the WERRs for the proposed models and the baselines on three test sets. By leveraging the post-AEC signals along with the AFE ouput, all the fusion network based CT models perform better than using AFE only (positive WERR values). Besides, the proposed NBF+F-LSTM and CNN fusion networks perform better than the channel-attention based fusion network, which implies better time vs. frequency relationship learning across channels given the comparable model size (Section~\ref{fusion_net_configs}). We can also see that the improvements from post-AECs and fusion network are more significant on \textit{MC Noisy} test sets than on \textit{MC Clean} test set. For example, the WERRs for NBF+FLSTM over AFE CT on \textit{MC Noisy} is $10.75\%$ versus $5.31\%$ on \textit{MC Clean}. It implies that the proposed methods are robust to challenging acoustic conditions.
\begin{figure}[t]
\centering
\includegraphics[scale=0.35]{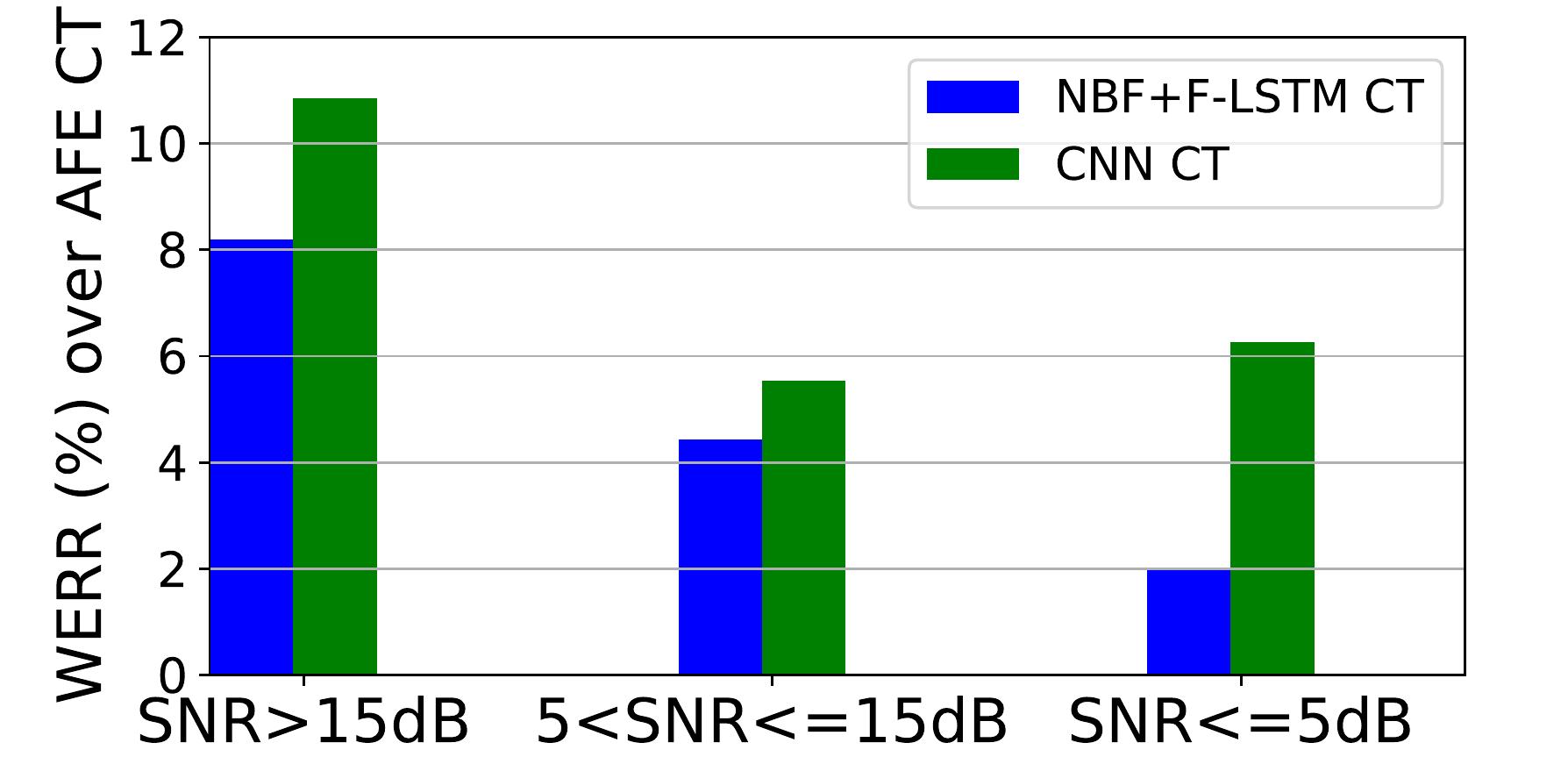}
\vspace{-0.2cm}
\caption{WERRs(\%) of the proposed methods \textbf{over AFE CT} w.r.t. different SNRs. A higher number indicates a better WER.}
\label{fig:werr_wrt_snr}
\end{figure}

\noindent \textbf{Impact of fusion networks and post AECs}: We evaluate the contribution of each sub-component in our models by removing either fusion networks (\textbf{no fusion net}) or post-AECs (\textbf{no post-AECs}). Note that \textbf{no fusion net} simply concatenates the AFE output and post-AECs and projects to the $d_{flstm}$/$d_{cnn}$=192-dim with a FFN layer. The results are illustrated in Table~\ref{tab:werrs_ablation}. 
For NBF+F-LSTM fusion based model, we see that removing post-AECs leads to much smaller WER improvements than removing the fusion network. In the CNN fusion based model, we see the similar trends on \textit{HEAD} and \textit{MC Noisy}. Adding post-AECs does not lead to further improvement (6.10\% vs. 8.66\% from AFE+CNN CT) on \textit{MC Clean} in this case.
\\ \textbf{Robustness to different SNRs}: In Fig.~\ref{fig:werr_wrt_snr}, we further evaluate the model robustness w.r.t. different SNR levels on the \textit{HEAD} test set.
It can be observed that both proposed methods offer improvements across the entire range of SNRs. In accordance to Table 1, the CNN fusion network offers the best improvements.
\\ \textbf{Robustness to device locations and noise types}: Finally, we evaluate our models
under different combinations of device locations and noise types provided in \textit{MC Noisy} test set. The results are presented in Table~\ref{tab:werr_device_locs_and_noises}. We see that the proposed models outperform the baselines with no post-AECs under \textbf{DevLoc2}, from $10.52\%$ (NBF+FLSTM CT under Silence noise) to $25.91\%$ (CNN CT under Appliance noise) relative WER improvement. This shows post-AECs benefits AFE  for the far-field ASR. 
Under \textbf{DevLoc1}, 
we observe smaller improvements with Silence noise as well as Appliance noise, yet $2.28\%$ and $6.71\%$ relative degradations under Media noise. We speculate that in DevLoc1 the noise source is much closer to the target speaker than in DevLoc2.
The post-AECs are highly contaminated by Media noise, as a result, do not provide useful complementary information to recover the errors produced by the AFE. In this case, including post-AECs could be detrimental.


\begin{table}[h]
    \caption{WERRs (\%) of the proposed methods \textbf{over AFE CT} w.r.t. different device locations and noise types. A higher number indicates a better WER.}
    \label{tab:werr_device_locs_and_noises}
    \centering
    \resizebox{0.99\linewidth}{!}{
    \begin{tabular}{r|c|c|c}
    \toprule
        Device Location & Noise Type & NBF+F-LSTM CT & CNN CT  \\ \midrule
        \textbf{DevLoc1} & Silence & 8.23 & \textbf{12.05}   \\ 
         $\sim 5$ feet away  & Appliance & \textbf{8.21} & 6.87  \\
        in front of the speaker & Media & -2.28 & -6.71 \\ \hline
        \textbf{DevLoc2} & Silence & 10.52 & \textbf{13.86}   \\ 
        $\sim 15$ feet away  & Appliance & 21.47 & \textbf{25.91}  \\
        in front of the speaker & Media & 12.76 & \textbf{17.92} 
    \end{tabular}}
\end{table}

\section{Conclusion}
We proposed two fusion networks based on NBF+F-LSTM and CNN in order to exploit the redundancy and aggregate the AFE output and post-AEC signals. The experimental results show that our fusion networks are better than the channel attention based fusion network given less than 2\% model parameter increase. In addition, our ablation studies demonstrate that exploiting both AFE and post-AEC mic array signals leads to the best improvement, e.g. 10.75\%  in \textit{MC Noisy} test set. Moreover, our methods are more robust to different SNRs, device locations, and noise types, achieving up to 25.91\% WER relative improvement over AFE CT.

\bibliographystyle{IEEEbib}
\small \bibliography{refs}

\end{document}